# Data-Driven Surrogate Modeling of DSMC Solutions Using Deep Neural Networks

Ehsan Roohi[*], Ahmad Shoja-sani

Mechanical and Industrial Engineering, University of Massachusetts Amherst,
160 Governors Dr., Amherst, MA 01003, USA. ([*]Corresponding author: roohie@umass.edu)

**Abstract**

This study presents a deep neural network (DNN) framework that accelerates Direct Simulation Monte Carlo (DSMC) computations for rarefied-gas flows, while maintaining high physical fidelity. First, a fully connected deep neural network is trained on high-quality DSMC data for seven temperatures (200 – 650 K) to reproduce the Maxwell–Boltzmann speed distribution of argon. Injecting the physical boundary point into the training set enforces the correct low-speed limit. It reduces the mean-squared error to below $10^{-5}$, thereby decreasing inference time from tens of minutes per DSMC run to milliseconds. For one-dimensional shock waves, a multi-output network equipped with learnable Fourier features learns the complete profiles of density, velocity, and temperature. Trained only on Mach numbers 1.4 – 1.9, it predicts a Mach 2 and 2.5 case with near-perfect agreement to DSMC, demonstrating robust out-of-training generalization. In a lid-driven cavity, the large parametric spread in Knudsen number is handled by a "family-of-experts" strategy: separate specialist models are trained at discrete Knudsen (Kn) values, and log-space interpolation fuses their outputs. This hybrid surrogate recovers the full 2-D velocity and temperature fields at unseen Kn with < 2 % spatial error. Key innovations include (i) explicit injection of physical constraints during data preprocessing, (ii) learnable Fourier feature mapping to capture steep shock gradients, and (iii) a modular expert-interpolation scheme to cover wide Knudsen ranges. Together, they establish a general recipe for trustworthy, rapid surrogate models that can be extended to non-equilibrium phenomena, gas mixtures, and design optimization workflows.

**Keywords:** Deep Neural Networks (DNN), Direct Simulation Monte Carlo (DSMC), Rarefied Gas Dynamics, Surrogate Modeling, Knudsen Number, Extrapolation, Fourier Feature Mapping, Specialist Models, Lid-Driven Cavity Flow

## 1. Introduction

Direct Simulation Monte Carlo (DSMC) is a particle method for simulating rarefied-gas flows in which the continuum Navier–Stokes equations break down. First formalized by G. A. Bird in the 1960s and codified in his widely cited monograph [1], DSMC separates molecular motion and intermolecular collisions over discrete time–steps that are smaller than the local mean collision time. Particles are streamed deterministically across a background grid, after which stochastic binary collisions are executed within each cell to reproduce the Boltzmann collision integral [2]. The algorithm's simplicity, modularity, and inherent parallelism have made it the reference tool for high Knudsen problems, ranging from hypersonic re-entry to micro-electro-mechanical systems (MEMS) [3].

Central to DSMC is the choice of collision model. Early implementations employed the Hard-Sphere (HS) model; however, the Variable-Hard-Sphere (VHS) and Variable-Soft-Sphere (VSS) models quickly became the standard because they replicate realistic viscosity–temperature power laws [1, 4]. Internal-energy exchange is typically handled using the Larsen–Borgnakke (LB) scheme, whereas modern codes incorporate chemical reactions and ionization through dedicated state-specific cross-sections. Efficiency improvements include the No-Time-Counter (NTC) procedure, the Majorant-Frequency Scheme (MFS) proposed by Ivanov and Markelov [5], and the Bernoulli Trial family [6-8]. Despite these advances, high-fidelity DSMC remains computationally intensive, motivating the use of surrogate approaches, such as the deep neural network framework presented in this work, to accelerate predictions without altering the underlying DSMC physics.

This paper details the development and validation of a deep neural network (DNN) designed to serve as a fast and accurate surrogate model for a fundamental DSMC simulation. Over the past years, a succession of studies has progressively shown that deep neural networks can capture the high-dimensional physics of molecular collisions and thus accelerate, or even supplant, classical Boltzmann solvers. Xiao & Frank [9] provided the first proof-of-concept by training a fully connected network to emulate the velocity-space collision integral for a hard-sphere gas; by enforcing mass, momentum, and energy conservation as soft constraints, their network reproduced both transient relaxation and steady-state Maxwellian statistics while delivering a two orders-of-magnitude speed-up on a CPU. Extending the idea, Roberts et al. [10] introduced a convolutional auto-encoder that learns a latent representation of post-collision velocity distributions; coupled with a "majorant frequency" sampling strategy, their NN–BGK operator ran nearly 10000 times faster than DSMC and maintained errors below 1 % for temperature and heat flux in homogeneous relaxation tests. Zhang et al. [11] subsequently embedded that operator into a hybrid lattice-Boltzmann framework, enabling continuum–rarefied simulations on uniform grids; their solver captured Knudsen-layer corrections around micro-pillars and achieved GPU runtimes comparable to standard LBM while remaining valid up to Kn ≈ 0.2. Parallel to these data-driven surrogates, Thapa & Karniadakis [12] developed a physics-informed neural network (PINN) that enforces the spatially homogeneous Boltzmann equation via automatic differentiation; using quadrature-free evaluation of the collision term, their PINN reproduced H-theorem decay rates without any labelled DSMC data, illustrating the potential of "label-free" training. Wu et al. [13] addressed the well-known spectral bias of multilayer perceptrons by equipping a PINN with *learnable* Fourier-feature encodings; the resulting model resolved Mach-10–20 normal shocks with a thickness error below 3 % and remained stable under mesh refinement, a feat unattainable by vanilla PINNs. Most recently, Tatsios and co-workers [14] proposed a hybrid continuum–particle scheme in which a DNN surrogate, calibrated on sparse DSMC snapshots, supplies closure terms to a Navier–Stokes solver for low-speed micro-flows; their Bayesian regression framework preserved confidence intervals and produced near real-time predictions on commodity GPUs.

In this work, we demonstrate that deep-learning surrogates can replace the DSMC simulations with high accuracy. We utilize a feed-forward network trained on a few equilibrium datasets to reproduce the Maxwell–Boltzmann velocity distribution function. A second, Fourier-enhanced model captures full shock wave profiles and extrapolates from training at M = 1.4–1.9 to M = 2–2.5 with negligible error. In addition, a "surrogate" scheme combined with logarithmic interpolation reconstructs lid-driven-cavity fields across two decades of Kn with < 2 % deviation. These advances—physically constrained training, learnable Fourier features, and modular Kn-wise experts—distinguish our framework from previous NN-Boltzmann efforts and provide a template for rapid, physics-faithful surrogates in DSMC simulations for non-equilibrium gas dynamics.

## 2. Machine Learning Strategy

### 2.1. Argon Relaxation to Equilibrium

#### 2.1.1 DSMC Simulations in Python

A surrogate model learns the complex input-output relationship of a high-fidelity simulation and provides nearly instantaneous predictions once trained. The first part of this paper utilizes a deep neural network (DNN) to develop a model for a classic DSMC problem: the thermal relaxation of a gas to its equilibrium state. The primary objective of this test is to develop a robust DNN surrogate model capable of accurately predicting the Maxwell-Boltzmann probability density function (PDF) for the speed of argon gas particles at thermal equilibrium. The model is designed to take speed (v) and temperature (T) as inputs and output the corresponding probability density P(v, T). The success of the model is evaluated based on its accuracy in comparison to theoretical solutions and its ability to generalize to temperatures not included in the training set.

The development process was divided into three key stages: data generation via DSMC, data preprocessing, and neural network training. A custom one-dimensional DSMC solver was implemented in Python to track the thermal relaxation of argon molecules inside an idealized, periodically repeating box. Although Python offers rapid prototyping, its default byte-code execution is too slow for a high number of particle moves and binary collisions required by DSMC. To bridge that gap, the computational hotspots—loops over particles, cell-based collision sampling, random number generation, and velocity updates—were annotated with Numba decorators.

Numba is an open-source just-in-time (JIT) compiler that translates a subset of Python and NumPy code into optimized machine code using the Low-Level Virtual Machine (LLVM) toolchain at runtime. When the DSMC script first executes, Numba inspects the function's argument types, specializes the abstract syntax tree, and emits LLVM Intermediate Representation (IR), which is then compiled to native instructions. The net result is a one-to-two-order-of-magnitude speed-up compared with an equivalent pure-Python implementation—sufficient to generate high-quality relaxation data for seven temperatures in a matter of minutes on a single workstation.

To ensure the model learned from accurate data, a high-quality dataset was generated using the following parameters: Particles per Cell: 500, Simulation Time: 8.0e-7 s (with sampling starting at 4.0e-7 s), Number of Bins for Histogram: 200. Simulations were executed for a range of seven distinct temperatures to provide a comprehensive training basis: [200.0, 275.0, 350.0, 425.0, 500.0, 575.0, 650.0] K. The collected particle speeds from these simulations formed our ground-truth dataset. For each temperature, the collected speeds were converted into a probability density function. The input features (X) were created as pairs of [bin center speed, temperature], and the target labels (y) were the corresponding probability densities from the histograms. A key insight during development was that the histogram data does not explicitly teach the model that the probability density must be zero at zero speed. To enforce this physical law, the data point (v=0, P=0) was manually added to the training set for each temperature. This simple addition dramatically improved the model's accuracy at low speeds.

The input data X was normalized using Standard Scaling: $X_{norm} = (X - X_{mean}) / X_{std}$, The output data y was normalized using Max Scaling: $y_{norm} = y / y_{max}$. This method was chosen over Standard Scaling to ensure that all target values remained non-negative (in the range [0, 1]), which is compatible with the final layer's activation function.

### 2.1.2. Neural Network Architecture and Training

A fully connected, feedforward deep neural network was constructed using TensorFlow and Keras. TensorFlow is an open-source numerical computing library developed by Google that organizes calculations as data-flow graphs: multidimensional arrays called tensors move along directed edges while nodes execute operations such as matrix multiplications, convolutions, or reductions. Keras is a high-level Application Programming Interface (API) that is now integrated into TensorFlow. It offers user-friendly abstractions—layers, loss functions, optimizers, callbacks—so building a deep neural network is reduced to stacking layers in a few lines of code. Under the hood, Keras objects create TensorFlow graphs, delegate tensor operations to the TensorFlow backend, and utilize its automatic differentiation and hardware acceleration capabilities.

The Architecture of the neural network, as shown in Fig. 1, is as follows: Input Layer: 2 neurons (for speed and temperature), Hidden Layers: Four hidden layers with 256, 256, 128, and 128 neurons, respectively. The ReLU (Rectified Linear Unit) activation function was used for all hidden layers, Output Layer: A single neuron with the softplus activation function. Softplus was chosen because it constrains the model's output to be non-negative, which is a physical requirement for a probability density.

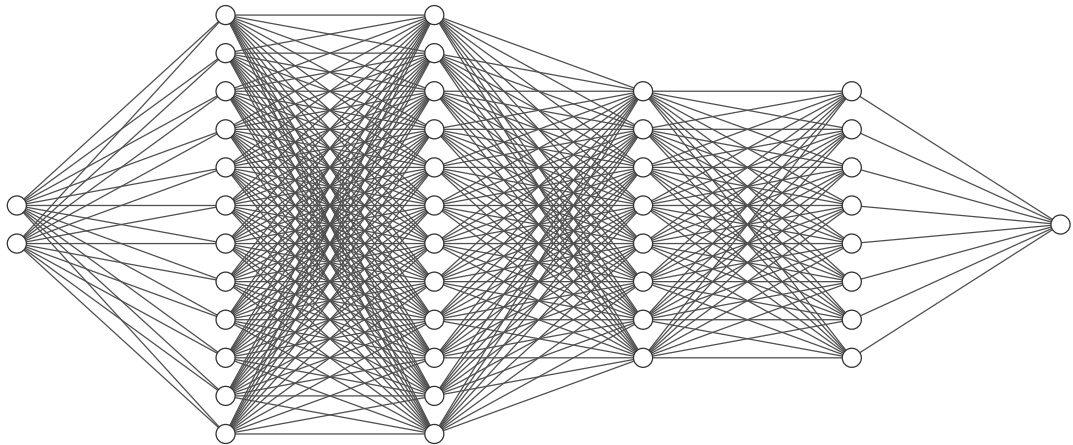

Fig. 1: The structure diagram of the employed neural network.

The training process is as follows: The Adam optimizer was used with an initial learning rate of 0.001. Mean Squared Error (MSE) was used to quantify the difference between the model's predictions and the true values. Two key callbacks were employed to manage the training: ReduceLROnPlateau automatically reduces the learning rate if the validation loss stops improving. EarlyStopping halts the training process if the validation loss does not improve for a set number of epochs (patience=75), preventing overfitting and saving time. The model was trained for a maximum of 250 epochs with a batch size of 64, using an 80/20 split for training and validation data. The training and validation loss curves converge and plateau together, as shown in Fig. 2. This behavior is the hallmark of a well-generalized model that has successfully learned the underlying signal without fitting the noise.

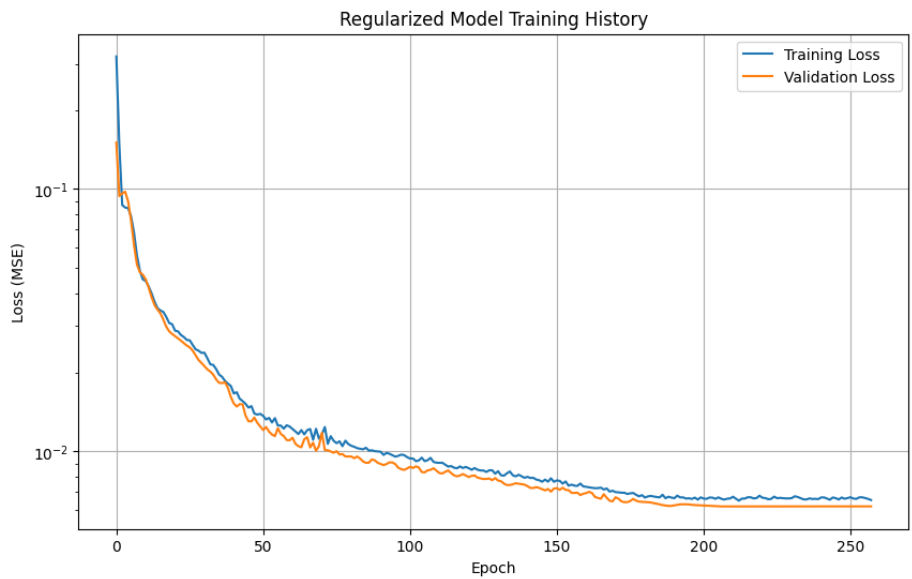

Fig. 2: The training and validation loss curves for the DNN model over 250 epochs.

The trained model demonstrated exceptional performance, successfully learning the underlying physics of the Maxwell-Boltzmann distribution. The model's predictions were compared against the theoretical Maxwell-Boltzmann PDF. For temperatures included in the training set (e.g., 500 K), the model's predictions exhibit an excellent match with both the theoretical curve and the raw DSMC data. As shown in Fig. 3, which depicts the VDF of the DNN, DSMC, and theory, the training phase was highly successful. The theoretical relation is given by:

$$P_v(v)dv = 4\pi\left(\frac{m}{2\pi kT}\right)^{3/2} v^2 e^{-0.5mv^2/kT} dv \tag{1}$$

The network learned the target distribution, demonstrating that the learning process was effective and well-tuned.

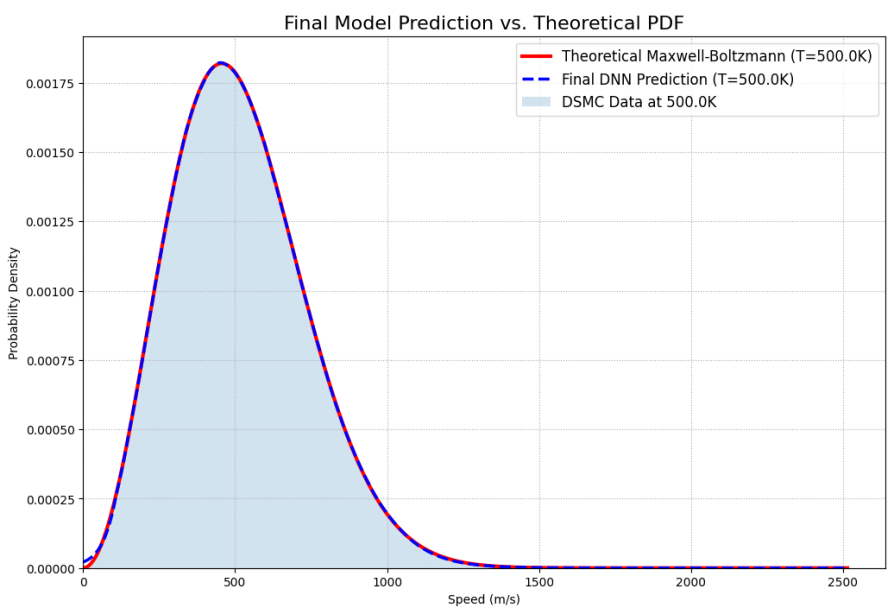

Fig. 3: Comparison of the theoretical Maxwell-Boltzmann analytical relation with the trained DNN at a temperature of 500 K.

### 2.1.3. Results

The most significant result is the model's ability to interpolate. When tested on a temperature it had never seen before (e.g., 325K), the model produced a highly accurate prediction that closely followed the theoretical curve. This confirms that the network did not simply "memorize" the data but learned the continuous relationship between temperature and the speed distribution. A primary goal of this test was to accelerate the simulation process. This was achieved decisively: Generating the high-quality data for a single temperature took upwards of a few minutes. However, once trained, the network can predict the entire distribution curve (for 300+ speed points) in milliseconds. This represents a speed-up factor, demonstrating the viability of this approach for applications requiring rapid or repeated evaluations.

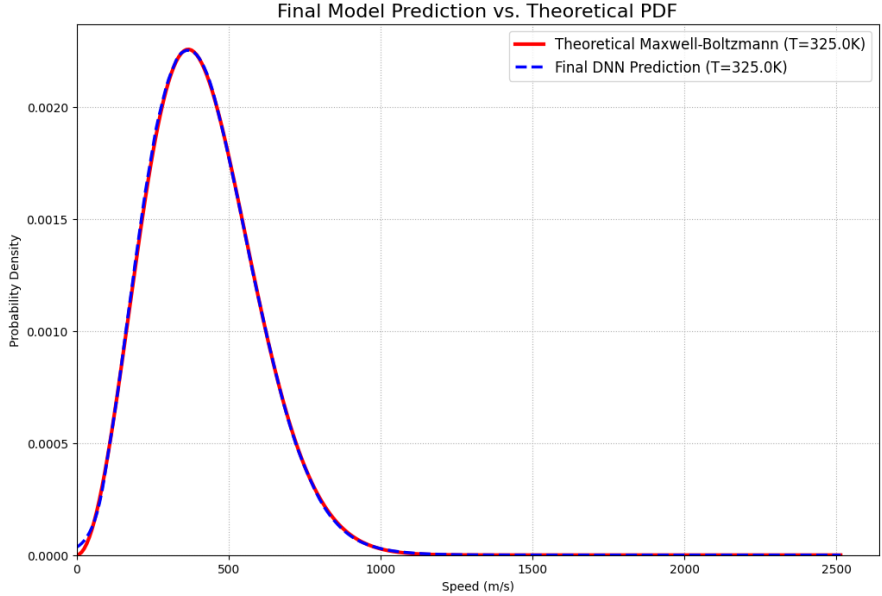

Fig. 4: Comparison of the theoretical Maxwell-Boltzmann analytical relation with the DNN prediction at an untrained temperature of 325 K.

This first test successfully developed and validated a deep neural network surrogate model for predicting the equilibrium speed distribution in a rarefied gas. By carefully preprocessing the data generated from DSMC simulations, particularly by incorporating a key physical boundary condition, the model achieved high accuracy and robust interpolation capabilities. The immense speed-up offered by the surrogate model opens the door to more complex computational studies that are currently intractable with DSMC alone.

### 2.1.4 Comparative Analysis with a Gradient Boosting Model

To benchmark the performance of the DNN, a comparative study was conducted against a state-of-the-art, non-neural-network model. We selected LightGBM, a high-performance gradient boosting framework, for this purpose. Gradient boosting models are ensembles of weak learners (decision trees). They are renowned for their exceptional speed and accuracy on structured, tabular data, making them a formidable baseline for this regression task.

### 2.1.4.1 Model Implementation: LightGBM

The LightGBM model was implemented using the official lightgbm Python library. The same high-quality, pre-processed dataset generated by the DSMC simulations was used for training. Unlike neural networks, tree-based models like LightGBM do not strictly require feature normalization. Therefore, the model was trained on the raw, un-normalized (speed, temperature) features to assess its performance under optimal conditions. The dataset was split into training (80%) and testing (20%) sets to ensure a fair evaluation on unseen data, as was done for the previous model. The model was configured with the following Hyperparameters: Regression $L_1$ (Mean Absolute Error, MAE). $L_1$ loss was chosen for its robustness and its tendency to be less sensitive to outliers compared to $L_2$ (MSE) loss. The number of estimators was set to a high value of 2000. To prevent overfitting and find the optimal number of trees, an early stopping callback was used. The training was configured to halt if the validation loss did not improve for 50 consecutive rounds. This ensures the model is both well-trained and efficient. A learning rate of 0.05 was used to ensure stable convergence.

**2.1.4.2 Results and Discussion**

The trained LightGBM model was evaluated both quantitatively and qualitatively. The model trained extremely quickly, completing in a matter of seconds. On the held-out test set, the trained LightGBM model achieved a final Mean Squared Error (MSE) of $1\times10^{-5}$. The predictive performance of the LightGBM model for an unlearned temperature of 325.0 K is shown in Figure 5.

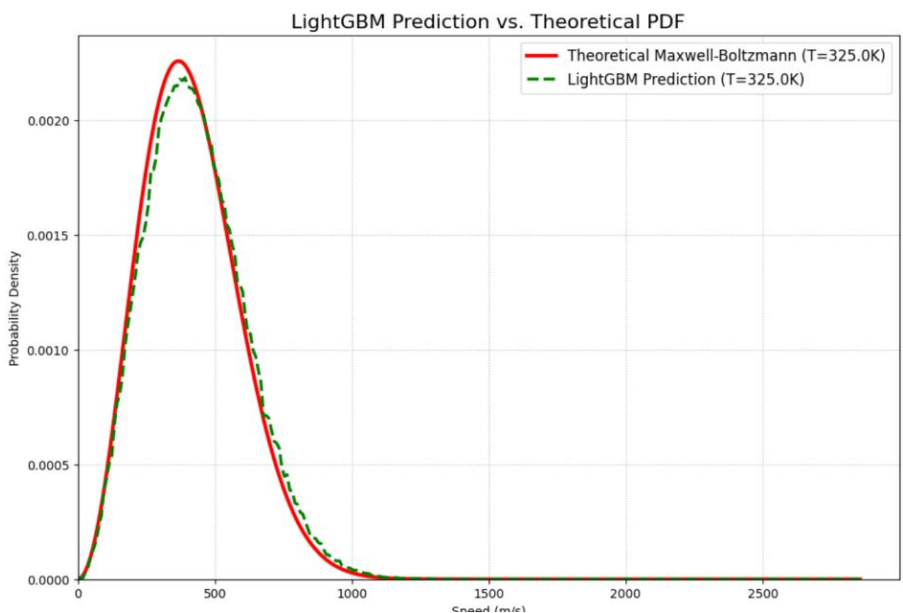

Fig. 5: VDF from LightGBM model for an unlearned temperature of 325.0 K compared to the analytical solution.

As can be seen, the LightGBM model successfully learned the general unimodal, bell-curve shape of the Maxwell-Boltzmann distribution. However, upon closer inspection, clear discrepancies with the theoretical curve are evident. The prediction (green dashed line) exhibits two primary inaccuracies: First, Peak Amplitude Error, where the predicted peak of the distribution is visibly lower than the theoretical peak. Second, Peak Location Error, where the location of the predicted peak is slightly shifted to the right (towards higher speeds) compared to the theoretical curve.

The comparative analysis reveals a clear trade-off between training speed and predictive fidelity. The DNN was demonstrably superior in terms of accuracy. Its predictions were virtually indistinguishable from the theoretical curve, correctly capturing the peak's height and location, as well as the physically-mandated behavior at zero speed. The LightGBM model, while capturing the overall trend, failed to replicate these finer details with the same level of precision. This difference is likely attributable to the fundamental nature of the models; DNNs are universal function approximators well-suited to learning smooth, continuous functions, whereas tree-based models create step-wise approximations that can struggle to perfectly represent such a function. On the other hands, LightGBM held a significant advantage in training time, converging orders of magnitude faster than the DNN. For applications where the highest possible fidelity and adherence to physical laws are paramount, the results strongly justify the selection of the DNN. The additional computational cost of training the DNN is a worthwhile investment to achieve a surrogate model with superior accuracy for this physics-based regression task.

### 2.1.4 Physics-Informed Neural Network (PINN) for Maxwell-Boltzmann Distribution

When we attempted to use the DNN discussed in Section 2.1.2 for unlearned temperatures outside of the learning range, i.e., extrapolation, the DNN demonstrated the limitations of a conventional DNN. We now propose a robust solution utilizing a Physics-Informed Neural Network (PINN) methodology. By incorporating physical scaling laws to non-dimensionalize the problem space, we transformed a difficult extrapolation task into a function approximation problem. The resulting physics-informed model accurately predicts the speed distribution for temperatures far outside the training range, showcasing superior accuracy, data efficiency, and physical consistency compared to the naive DNN approach.

The training data for the models was generated via a high-fidelity DSMC simulation for Argon gas at several equilibrium temperatures ranging from 200 K to 650 K. The Maxwell-Boltzmann distribution's dependence on temperature is well-defined. Here, we suggest defining a characteristic speed, $v_{char}$, that scales with temperature: $v_{char}=mk_BT$ where $k_B$ is the Boltzmann constant and m is the particle mass. Using this, we define a dimensionless speed, s: $s=v_{char}/v$. By performing a change of variables, the probability density function P(v) can be transformed into a dimensionless probability density P(s), where *P(s)ds = P(v)dv*. The crucial insight is that the resulting dimensionless distribution, $P(s)=2/\pi \times s \times \exp(-s/2)$, is a universal function that is independent of temperature. All DSMC data, regardless of the initial temperature, will collapse onto this single curve after non-dimensionalization.

### 2.1.4.1 PINN Model Architecture

The learning task for the network is now drastically simplified. A single scalar value, the dimensionless speed (*s*), is considered as input. The corresponding dimensionless probability density, P(s) is set as output. The same Multi-Layer Perceptron (MLP) architecture as the standard DNN was used, but its input layer was modified to accept a single-dimensional vector. The network is no longer required to learn the complex relationship with temperature; it only needs to approximate a single, universal 1D curve.

**2.1.4.2 Results and Discussion**

The model was evaluated on its ability to perform both interpolations, i.e., predicting for T=325 K, within the training range, and extrapolation, i.e., predicting for T=900 K, far outside the training range. The PINN model demonstrated exceptional performance in both scenarios.

Fig. 6 illustrates the training history of the employed DNN, showing the evolution of training and validation loss over 175 epochs. Both losses, measured in mean squared error (MSE), decrease rapidly during the initial epochs, indicating effective learning. From approximately epoch 50 to 140, the training and validation losses remain low and closely aligned, suggesting that the model generalizes well without significant overfitting. However, after epoch 150, both curves exhibit slight fluctuations, and the validation loss increases marginally, which may point to the onset of overfitting or sensitivity to learning rate adjustments. Overall, the training appears successful, with low final loss values and strong consistency between training and validation performance.

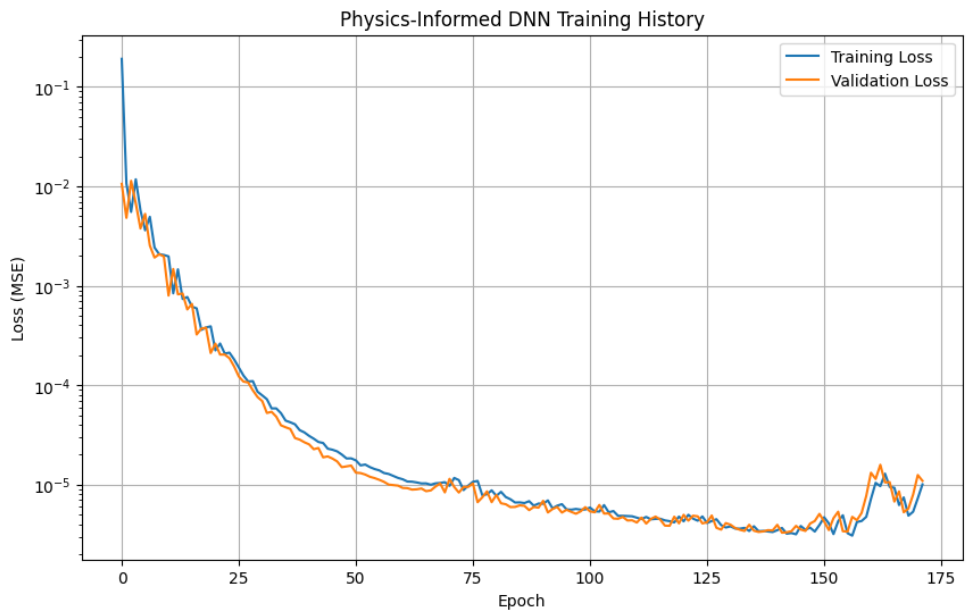

Fig. 6: The training and validation loss curves for the DNN model over 175 epochs.

As shown in Fig. 7, the model's predictions for unseen temperatures are nearly indistinguishable from the theoretical Maxwell-Boltzmann PDF. This success is directly attributable to the physics-informed approach. The challenging task of extrapolation in the temperature domain was not handled by the neural network, but rather by the analytical scaling laws applied during the pre-processing and post-processing steps. The neural network was responsible only for the simple task of interpolating along the universal dimensionless curve, which it accomplished with high fidelity.

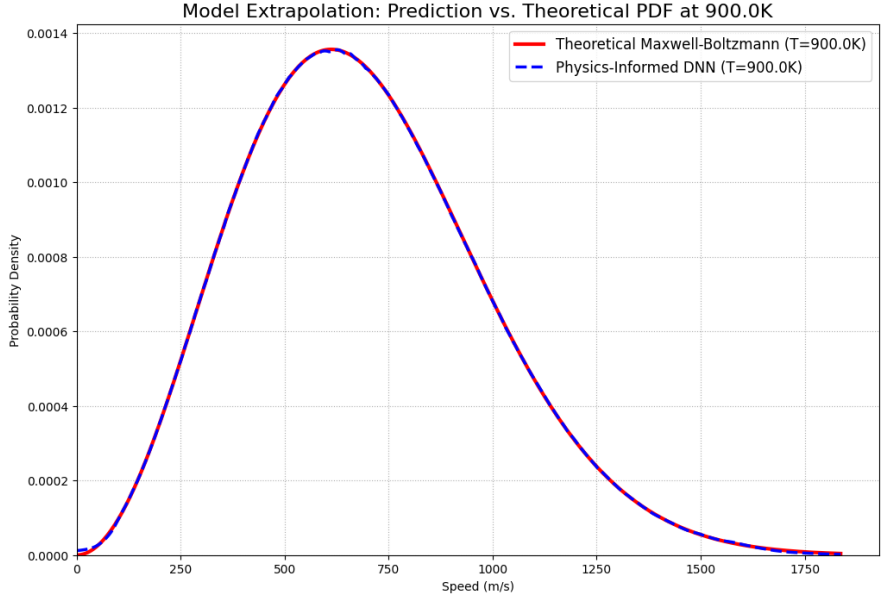

Fig. 7: Comparison of the theoretical Maxwell-Boltzmann analytical relation with the DNN prediction at an untrained temperature of 900 K, i.e., extrapolation.

**2.2 Deep Neural Network for Shock Wave Profile Prediction**

Here, we focus on the development and evaluation of a DNN designed to predict the physical properties of a one-dimensional shock wave. The objective is to create a model capable of learning the complex, non-linear relationships between the spatial position across a shock and its resulting fluid dynamic properties. The model was trained on data generated from DSMC for various upstream Mach numbers. The DSMC simulations were performed using the code DSMC1S.FOR of Bird, which models the structure of a one-dimensional shock wave using 300 physical cells (MNC=300), each divided into 6 sub-cells (NSC=6), resulting in a total of 1800 sub-cells. The upstream boundary is set to negative and the downstream boundary to positive, with a fraction of the domain upstream. The upstream flow is defined by velocity, temperature, and number density, with the shock Mach number computed from these values. The time step for molecular movement and collision is constant, with sampling performed every NIS steps and printed every NSP samples until reaching a total of NPT output intervals. The simulation includes a single gas species, Argon. The molecular diameter of argon is given as $4.092 \times 10^{-10}$ meters, with a reference temperature of 293 K. The viscosity-temperature index is 0.81. The reciprocal of the variable soft sphere (VSS) scattering parameter is set to 0.6015. The molecular mass is $6.64 \times 10^{-26}$. We used a total of up to 20,000 simulated molecules. The ultimate goal is to create a surrogate model that can rapidly predict shock profiles without running DSMC simulations.

**2.2.1 Dataset**

The dataset consists of multiple text files, each corresponding to a DSMC simulation at a specific Mach number. Each file contains tabular data with six columns: Position (x), Density (ρ), Velocity (u), Temperature (T), and Translational Temperature ($T_{trans}$). The model was trained on data from Mach numbers 1.4, 1.5, 1.6, 1.8, 1.9, and 2.0. The data for Mach 1.7 was held out and used exclusively as the test set to evaluate the model's ability to predict a profile within the bounds of its training data.

### 2.2.2 Neural Network Architecture

A feedforward deep neural network was constructed using the TensorFlow and Keras libraries. The model takes two features as input—Position and Mach Number—and simultaneously predicts the four corresponding physical properties. The neural network architecture, shown in Fig. 8, consists of an input layer that accepts a vector of two features—Position and Mach Number—followed by three hidden layers, each comprising 64 neurons with ReLU activation functions. The final output layer contains 4 neurons with a linear activation function, corresponding to the predicted physical properties: density, velocity, and temperature.

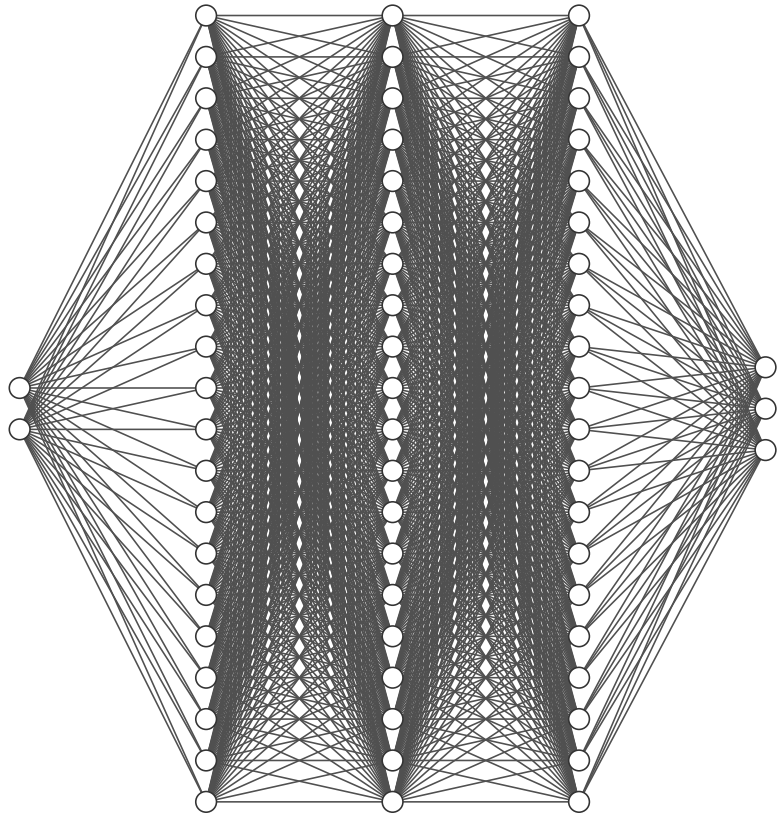

Fig. 8: The structure diagram of the employed neural network for the normal shock problem.

### 2.2.3. Training and Preprocessing

Both input and output features were normalized to a range of [0, 1] using MinMaxScaler from scikit-learn. This is a critical step that helps stabilize training and improve model convergence.

The scaler was fitted on the training data only and then used to transform the test data. The model was compiled with the Adaptive Moment Estimation (Adam) optimizer and the Mean Squared Error (MSE) loss function. It was trained for 200 epochs with a 20% validation split to monitor for overfitting. The Adam optimizer is a widely used algorithm for training deep neural networks, known for its efficiency, robustness, and fast convergence. It combines the benefits of both AdaGrad and RMSProp by maintaining adaptive learning rates for each parameter and computing exponentially decaying averages of past gradients (first moment) and squared gradients (second moment).

### 2.2.4 Results and Discussion

The trained model was evaluated on the unseen Mach 1.7 test case. The results, shown in Fig. 9, demonstrate a strong predictive capability, as the model successfully captures the shock wave's structure for all three output variables compared with the DSMC, referred to here as the exact solution. The model accurately predicted both the pre-shock and post-shock conditions, and, most importantly, the steep gradients within the shock layer itself. This indicates that the network effectively learned the underlying physics of the non-equilibrium zone.

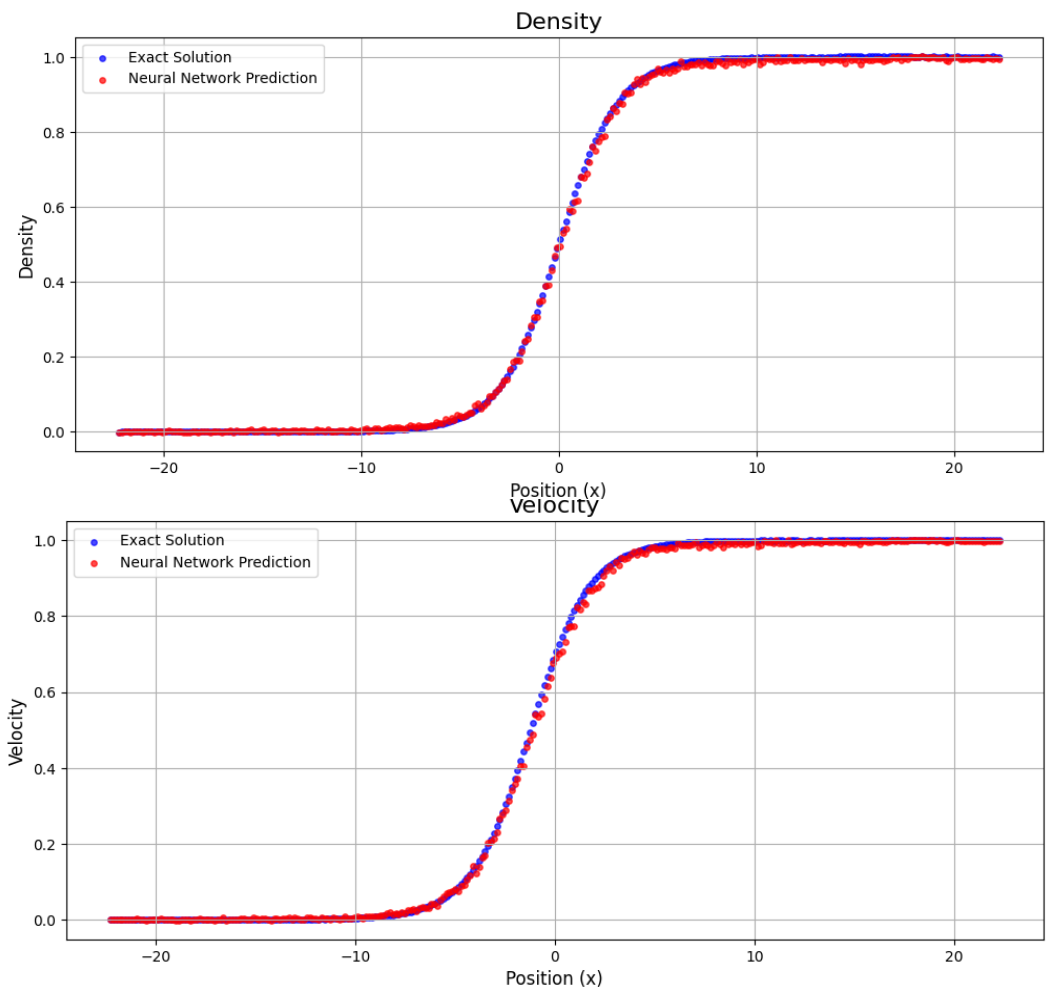

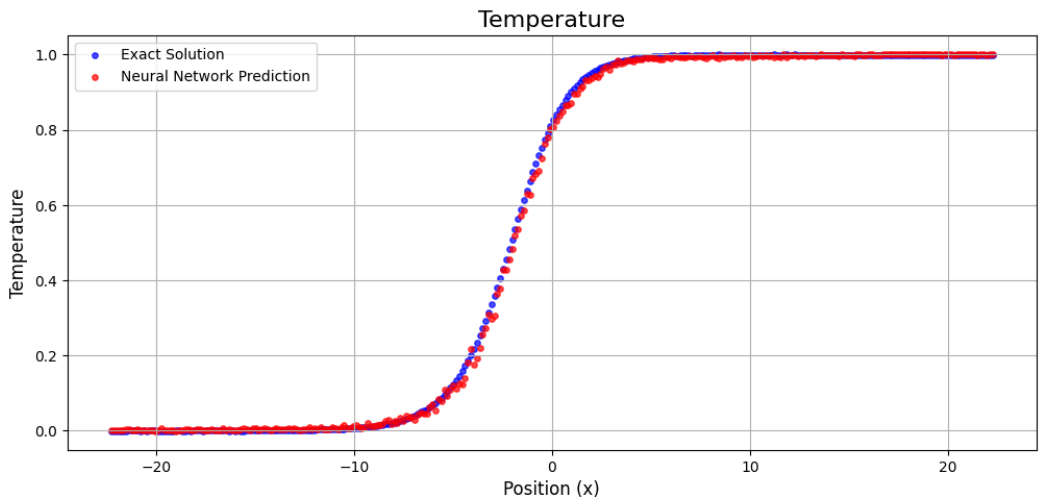

Fig. 9: DNN prediction of the density, velocity, and temperature for the shock wave at an unlearned Mach number of 1.7.

### 2.2.5 An Optimized Neural Network for Shock Wave Extrapolation

Here, we aim to develop a DNN capable of accurately extrapolating the velocity profile of the one-dimensional shock wave. While the standard neural networks, like the previous one, often excel at interpolation (predicting within the domain of training data), they notoriously struggle with extrapolation (predicting outside the domain of learning). After extensive experimentation with various architectures and training techniques, we developed a final and highly successful model— a highly tuned, data-driven deep neural network. The key components of this successful strategy are detailed below.

### 2.2.5.1 Fourier Feature Mapping

A critical breakthrough in model accuracy was the application of Fourier Feature Mapping to the spatial coordinate input (x). Instead of feeding the raw position data to the network, the 1D position vector was mapped into a 128-dimensional feature space using a bank of sinusoidal functions. A mapping matrix B was randomly initialized from a normal distribution. The input x was transformed into a feature vector consisting of $[\sin(2\pi Bx), \cos(2\pi Bx)]$. This technique enables the network to learn high-frequency functions, such as the steep gradient of a shock wave, far more effectively. The network can construct the S-shaped shock profile by learning a linear combination of these sinusoidal basis functions, rather than trying to approximate it with piecewise-linear ReLU activations. This proved essential for capturing the sharp details of the shock.

Recent studies have emphasized and validated the effectiveness of Fourier feature mapping in neural networks. Tancik et al. [15] formally demonstrated that Fourier features transform the neural tangent kernel into a stationary kernel, enabling MLPs to learn high-frequency functions in low-dimensional domains efficiently. Later, Sen Li et al. [16] applied this concept in their F-D3M method, using Fourier features within a domain decomposition framework to solve partial differential equations with enhanced accuracy and convergence. More recently, Mema et al. [17] incorporated Fourier feature mapping into a Deep Ritz method to solve variational problems related to microstructure modeling, showing significant improvements in PDE solutions. Building upon these developments, Feng et al. [18] introduced SASNet, a spatially adaptive sinusoidal neural network architecture capable of reconstructing sharp implicit representations, further confirming the power of frequency-aware embeddings in capturing steep gradients and fine-scale features. These advancements collectively underscore the critical role of Fourier feature mapping in accurately modeling complex physical structures, such as shock profiles, by allowing neural networks to construct S-shaped transitions through linear combinations of sinusoidal basis functions instead of relying on piecewise-linear approximations

**2.2.5.2 Network Architecture**

The final architecture, shown in Fig. 10, was carefully designed to strike a balance between sufficient learning capacity and stable extrapolation behavior. The input layer accepts a 129-dimensional vector, consisting of 128 Fourier-mapped spatial features combined with a single Mach number feature. The network includes three fully connected hidden layers, each containing 128 neurons and using the hyperbolic tangent (tanh) activation function. This choice of activation was a crucial tuning step, as the smooth and nonlinear nature of tanh outperformed ReLU in modeling the continuous, physical nature of shock-wave structures, resulting in more accurate and smoother predictions. The output layer consists of a single neuron with a sigmoid activation function, which was essential for stable extrapolation. By constraining the output to the normalized range of [0, 1], the sigmoid activation eliminated the overshoot problem seen in earlier models, ensuring the network's predictions remained physically consistent.

The model was trained using the Adam optimizer, with a carefully selected learning rate of 0.0005. This slightly reduced rate, compared to the standard 0.001, enabled more precise convergence during the later stages of training without causing the optimization to stall. To ensure complete convergence at this lower learning rate, the network was trained for a total of 400 epochs. For preprocessing, the input features were normalized using a MinMaxScaler to ensure consistent scaling across the high-dimensional input space. The target variable, velocity, was not explicitly scaled, as the sigmoid activation in the output layer effectively constrained the predictions to the normalized range of [0, 1], making additional scaling unnecessary.

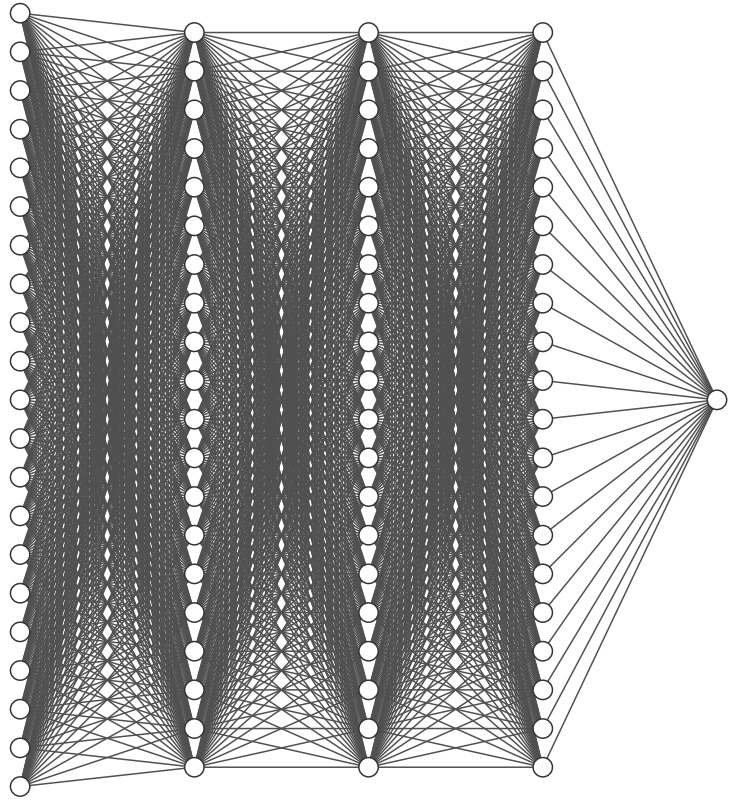

Fig. 10: The structure diagram of the employed neural network based on Fourier Feature Mapping for shock wave extrapolation.

### 2.2.5.3 Results for M=2.0

The final model demonstrated excellent performance on the Mach 2.0 extrapolation task, as shown in Fig. 10, where the predicted profiles for density, velocity, and temperature align remarkably well with the exact DSMC solution, despite the model being trained only up to Mach 1.9. This high-fidelity prediction validates the robustness and generalization capability of the developed surrogate model. Achieving such accuracy in a challenging extrapolation setting required an iterative tuning process and the integration of several advanced techniques: constraining the output space with a sigmoid activation function effectively eliminated boundary condition violations; enriching the input space using Fourier Feature Mapping enabled the network to capture the sharp gradients associated with shock structures accurately; and using the smooth tanh activation function in the hidden layers, coupled with a carefully reduced learning rate, allowed the high-capacity network to converge to a stable and precise solution. These elements collectively formed a reliable and physically consistent architecture capable of extrapolating conditions beyond its training domain. Nevertheless, to further extend the model's extrapolation capabilities to even higher Mach numbers, additional strategies are necessary, as discussed in the following section.

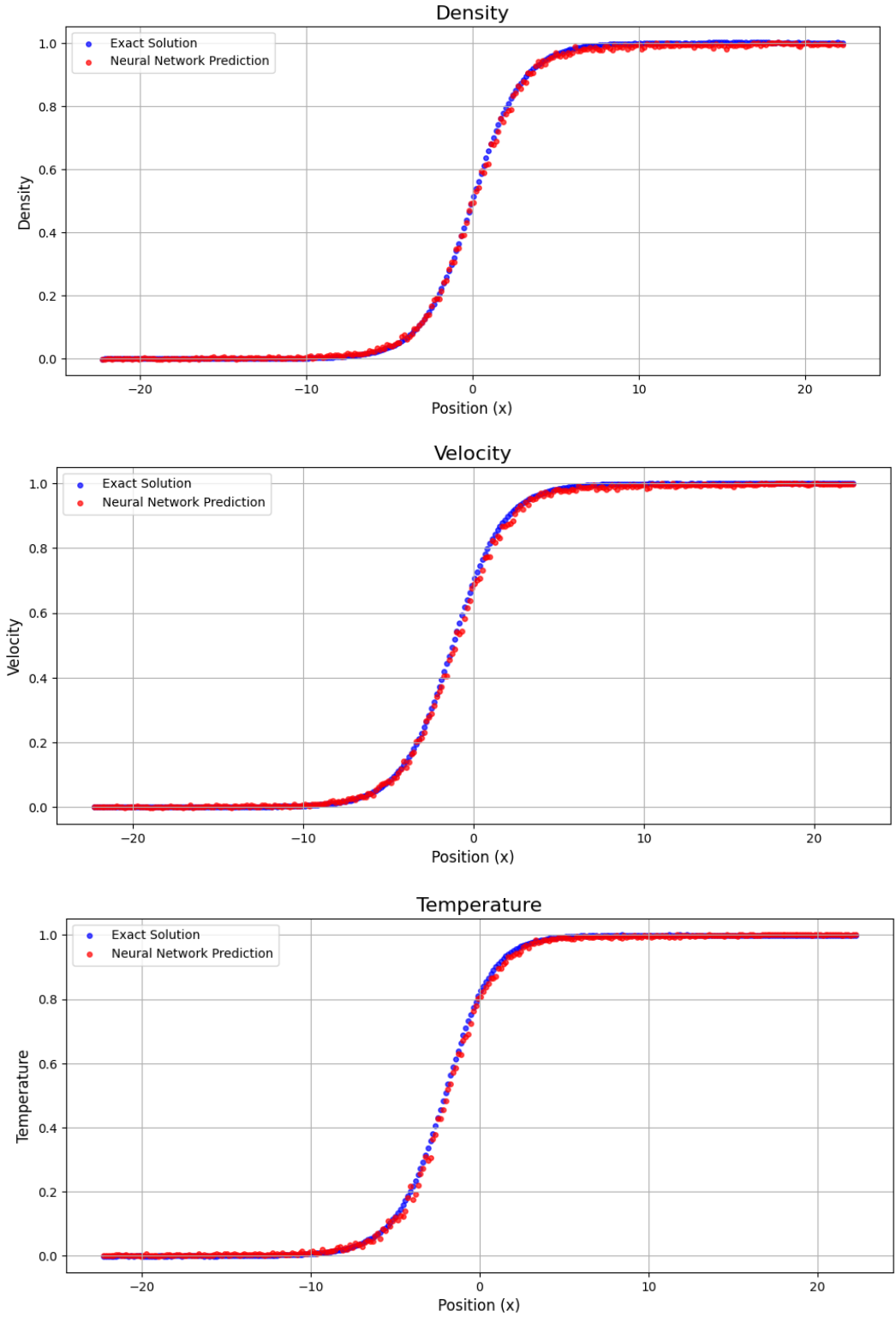

Fig. 10: DNN prediction of the density, velocity, and temperature for the shock wave at an unlearned Mach number of 2.

**2.2.5.4 Improvement in DNN for Extension to M=2.5**

The previous DNN architectures proved inadequate for predicting flow properties at Mach 2.5 when trained only on data up to Mach 1.9, as they failed to generalize beyond the training domain. We employed an iterative development process, exploring advanced architectures that included learnable Fourier Features and modulated networks. While these models offered increased expressive power, they introduced instabilities such as high-frequency oscillations and spatial phase shifts. The critical breakthrough was achieved by combining a robust concatenated architecture with strong regularization, specifically Dropout and L2 regularization, which stabilized the learnable Fourier Features. The final model's performance was found to be sensitive to hyperparameter tuning and the inherent stochasticity of the training process. By setting a random seed to ensure reproducibility, a definitive model was established that accurately predicts the shock profile, including its position, amplitude, and slope, demonstrating a successful methodology for robust extrapolation in complex physical systems.

Initial attempts to model shock structures using simple MLPs failed to capture the steep gradients, prompting the introduction of Fourier Feature positional encoding to represent high-frequency spatial variations better. However, advanced architectures based on this concept, such as models with learnable Fourier frequencies or Mach-modulated positional streams, proved unstable during extrapolation, leading to high-frequency noise or spatial phase shifts. These failures highlighted that increasing complexity alone was insufficient and often counterproductive. The breakthrough came by simplifying the architecture: a deep network that concatenated position and Mach inputs, paired with strong regularization techniques. Specifically, dropout layers (rate 0.3) were added after each dense layer to reduce overfitting, and $L_2$ regularization ($\lambda = 10^{-6}$) was applied to limit weight magnitudes. This approach stabilized training, suppressed oscillations, and yielded the first accurate extrapolation result, demonstrating that model regularization, rather than complexity, was crucial for generalizing beyond the training regime.

The final model's performance was found to be highly sensitive to the scale hyperparameter of the Fourier Features layer, which dictates the initial distribution of frequencies. Due to the stochastic nature of weight initialization, dropout, and data shuffling, repeated training runs yielded slightly different results. To ensure the final model is deterministic and its excellent performance is reproducible, a global random seed was set for all libraries (TensorFlow, NumPy, and Python's random) at the beginning of the script.

**2.2.5.5 Results for M=2.5**

The final model architecture integrates separate inputs for spatial position and Mach number, applies a learnable Fourier Feature mapping (with scale 5.0) to the position input, and concatenates this with the scaled Mach input before passing it through a deep stack of four dense layers with swish activations, each regularized using L2 penalties and a 30% dropout rate. A final sigmoid output layer ensures consistency with the normalized target velocity. As illustrated in Fig. 11, this architecture suitably predicts the velocity profile at Mach 2.5, correctly matching the DSMC reference in terms of shock position, amplitude, and slope, despite being trained on lower Mach data up to 1.9. This outcome underscores that achieving stable and accurate extrapolation in complex physical systems is not simply a matter of architectural complexity. Rather, it requires a carefully engineered combination of learnable representations, strong regularization, and tuned hyperparameters.

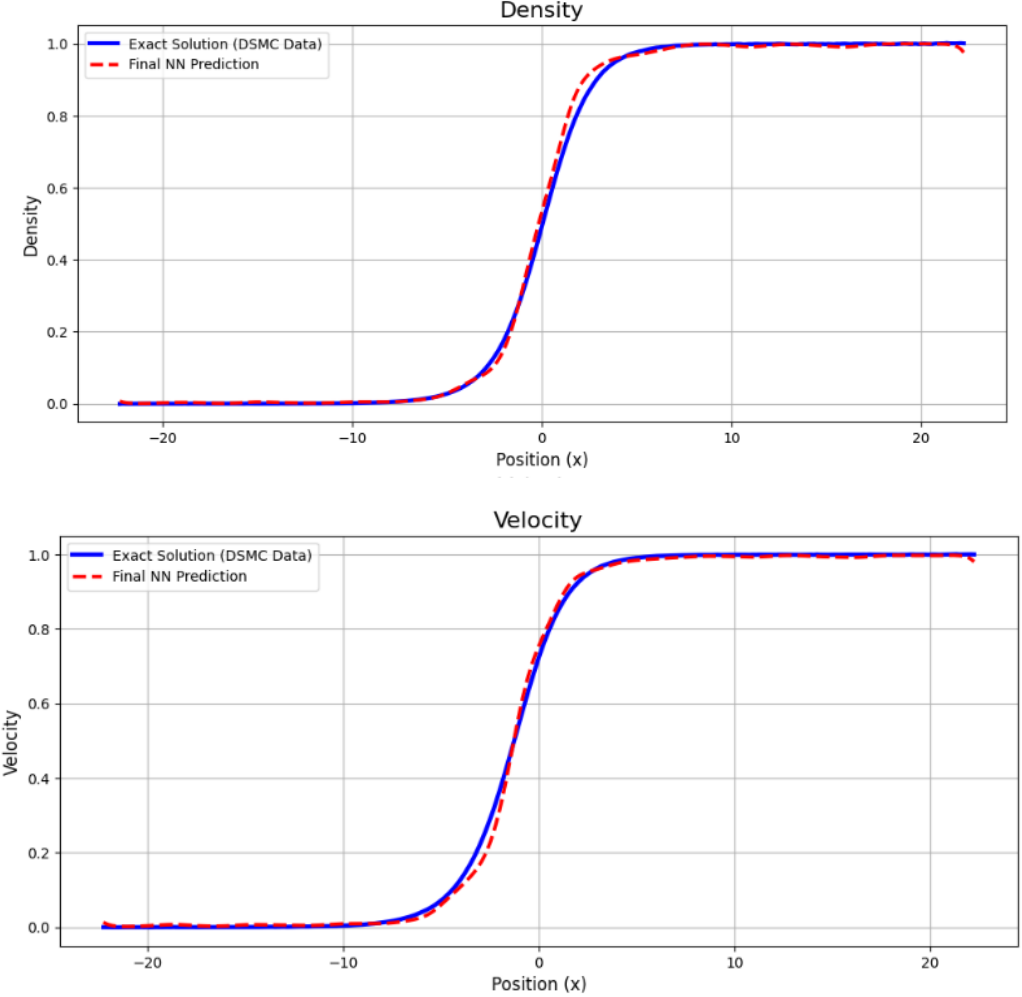

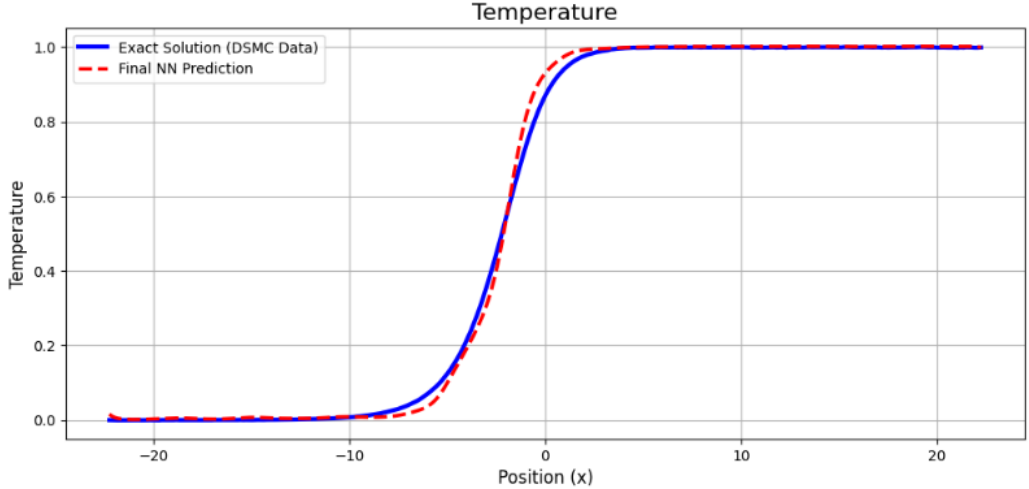

Fig. 11: DNN prediction of the density, velocity, and temperature for the shock wave at an unlearned Mach number of 2.5.

**2.3 A DNN Surrogate Model for Cavity Flow Predictions**

This section describes the development of a DNN to serve as an accurate surrogate model for predicting the flow field in a rarefied lid-driven cavity problem, which was originally solved using the DSMC method for a wide range of Knudsen numbers, i.e., Kn from 0.001 to 10. A "Hybrid Surrogate Model" approach was used.

**2.3.1 Problem Statement**

Lid-driven microcavity is considered an excellent benchmark problem for DSMC simulations [19-21]. Here, the cavity flow is simulated using Bird's standard no-time-counter (NTC) algorithm using a modified Bird DSMC2D.FOR code. Argon gas is used in the simulations, characterized by a molecular mass of $6.64 \times 10^{-26}$ kg and a molecular diameter of $4.092 \times 10^{-10}$ m. The gas is confined within a square microcavity with a side length $1 \times 10^{-6}$ m. Figure 12 illustrates the geometry and flow conditions of the test case. The lid of the cavity moves at a constant velocity of 100 m/s, and the Knudsen number transitions through various regimes, from slip to the free molecular regime. The cavity walls are maintained at non-uniform temperatures. Specifically, the lower corners of the cavity (points A and D) are set to 350 K. In comparison, the upper corners (points B and C) are maintained at 300 K. This configuration results in a uniform temperature of 300 K along the top (lid) wall and 350 K along the bottom wall. Meanwhile, the temperatures of the left and right sidewalls vary linearly from 300 K, near the top (lid), to 350 K near the bottom, creating a temperature gradient across the vertical boundaries.

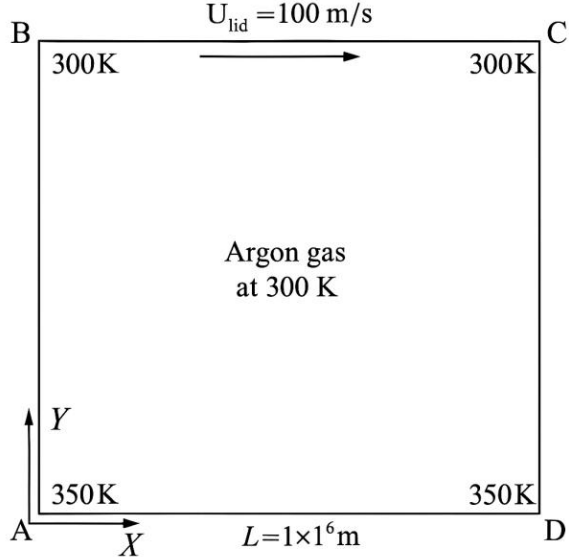

Fig. 12: The schematic of the cavity considered for DSMC simulations and DNN predictions.

**2.3.2 Modeling Strategy: A Hybrid Surrogate Model for Parametric DSMC Simulations of Cavity Flow**

This section outlines a successful methodology for creating a high-fidelity surrogate model for DSMC data of a lid-driven cavity flow across a range of Knudsen numbers. The approach is a hybrid, two-stage approach: first, training a "family of specialist models," where each neural network is an expert for a single, fixed Knudsen number. To the best of our knowledge, this strategy has not been applied to the fluid mechanics of rarefied gases so far. Second, employing a logarithmic interpolation scheme to synthesize predictions for any intermediate Knudsen number not seen during training. This method proved highly effective, accurately reproducing the complex 2D velocity and temperature fields, including the vortex structure.

The core challenge of this investigation is the significant variation in flow physics across the studied Knudsen numbers, i.e., Kn from 0.001 to 10, coupled with the sparsity of data points along this parametric dimension. A single neural network struggled to learn a generalizable function $f(X, Y, Kn)$.

To address the challenge of modeling across a wide range of parameters, we adopted a family of specialized surrogate models. The core methodology involved decomposing the complex, multi-Knudsen-number problem into a set of simpler, non-parametric subproblems. For each Knudsen number in the training set (e.g., Kn = 0.001, 0.01, 0.1, 1, 10), a dedicated high-fidelity neural network was trained to learn the spatial mapping specific to that regime. These specialist models captured detailed flow behavior at their respective Kn values. To predict flow fields at unseen Knudsen numbers (e.g., Kn = 0.05 in the first attempt and Kn=0.5 in the second attempt), we employed a synthesis step based on interpolation: predictions from neighboring specialist models were combined to approximate the desired solution. This strategy effectively transformed the challenging problem of extrapolation across physical regimes into a series of well-behaved spatial interpolation tasks, followed by a final interpolation in the parametric dimension, yielding robust and accurate predictions.

### 2.3.3 Specialist Model Architecture

#### 2.3.3.1 Feature Engineering: Positional Fourier Features

To enable the network to capture fine spatial details and smooth gradients, the 2D spatial coordinates (X, Y) were first transformed using a Fourier Feature layer. This layer maps the low-dimensional coordinates to a high-dimensional feature space of sine and cosine functions of varying frequencies:

$$\gamma(v) = [\cos(2\pi \times B \times v), \sin(2\pi \times B \times v)]$$

, where $v$ is the input coordinate vector and $B$ is a fixed, randomly generated frequency matrix. The scale hyperparameter of the random distribution for $B$ was tuned to 2.5 to provide a rich basis of frequencies suitable for resolving the flow structures without overfitting to high-frequency noise. The frequency matrix $B$ was kept non-trainable to ensure model stability.

#### 2.3.3.2 Network Body and Regularization

The high-dimensional feature vector from the Fourier layer was processed by a deep network consisting of three hidden dense layers, each with 256 neurons and using the swish activation function for efficient gradient flow (see the schematic in Fig. 13). Regularization was critical for achieving a smooth, physically plausible output. Dropout with a rate of 0.2 was applied after the first two hidden layers, and L2 regularization ($\lambda=10^{-6}$) was applied to all dense layer kernels. The final output layer employed a linear activation function, a standard choice for regression tasks, enabling the network to predict any real-valued number (positive or negative).

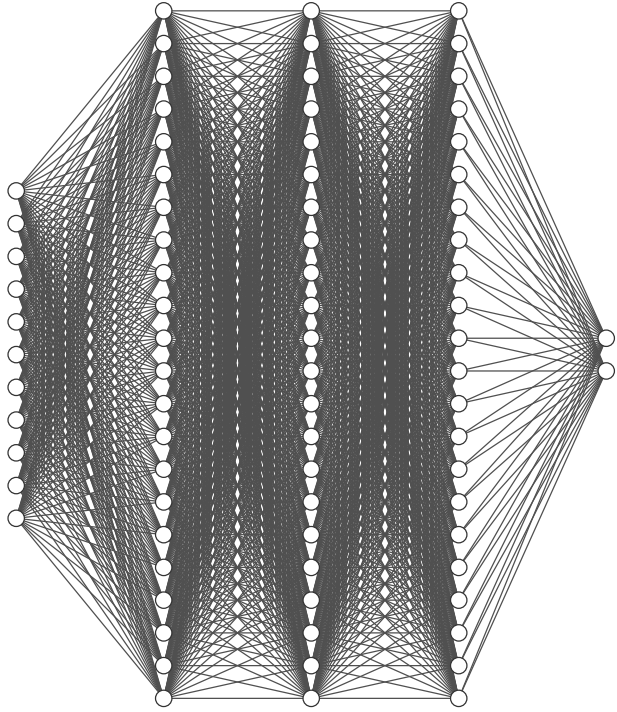

Fig. 13: The structure diagram of the employed neural network for the cavity problem.

**2.3.3.3 Data Preprocessing and Training Protocol**

For each specialist model in the family, the data corresponding to its specific Knudsen number was first shuffled and partitioned into a training set (85%) and a validation set (15%) using sklearn.model_selection.train_test_split, ensuring that validation loss accurately reflected generalization within that physical regime. The model inputs were defined as the spatial coordinates, while the outputs consisted of a vector containing all other physical variables available in the dataset, including velocity and temperature. Input coordinates were scaled to the [0,1] range using MinMaxScaler, while the output variables were normalized using StandardScaler, which centers data around a mean of 0 and standard deviation of 1, providing stability for variables with both positive and negative values, such as velocity components, and avoiding the limitations imposed by strict bounding. Each model was trained using the Adam optimizer with the mean_squared_error loss function, and both EarlyStopping and ReduceLROnPlateau callbacks were employed to ensure optimal convergence without overfitting.

**2.3.3.4 Interpolation Framework for New Predictions**

To predict the flow field at the unseen Knudsen number $Kn_{test}$=0.05, a hybrid interpolation strategy was employed using the trained specialist surrogate models. First, the two models corresponding to the closest Knudsen numbers bracketing the test value were identified. The spatial grid coordinates of the test case were then input into both models to generate two complete, high-fidelity flow field predictions. Recognizing that Knudsen numbers span multiple orders of magnitude, a logarithmic interpolation scheme was used to ensure physical consistency. A weight *w* was computed as:

$$w = \frac{\log_{10}(\text{Kn}_{\text{test}}) - \log_{10}(\text{Kn}_{\text{lower}})}{\log_{10}(\text{Kn}_{\text{upper}}) - \log_{10}(\text{Kn}_{\text{lower}})} \tag{2}$$

This methodology yielded an accurate and physically consistent prediction for a Knudsen number not included in the training set, demonstrating the effectiveness of the specialist-surrogate interpolation framework.

### 2.3.4 Results for the cavity flow

Fig. 14 presents a qualitative comparison between DSMC and interpolated DNN predictions for cavity flow at Kn=0.05. The top row displays velocity contour fields and streamlines from both methods: the left plot shows the DSMC result, while the right plot shows the DNN prediction. Both demonstrate a strong primary vortex with highly similar structure and magnitude, indicating excellent agreement between the surrogate and reference solutions. The bottom row compares the corresponding temperature fields, revealing nearly identical thermal gradients from the bottom to the top walls, which validates the DNN's ability to capture complex thermal behavior across the cavity. Note that the DSMC solution of this case takes around 5 hours, while the DNN prediction was performed on the same system within 10 minutes. Fig. 15 provides a quantitative comparison through line plots. The top plot shows U-velocity profiles along the x-direction at a horizontal slice (y = 0.8), comparing DSMC (solid lines) and DNN (dashed lines). The predictions align closely across all profiles, with minor deviations in peak velocities. The bottom plot presents the temperature distribution at the same slice, again showing strong agreement between the DSMC and DNN outputs, with only slight differences, which are negligible in magnitude. Overall, both figures demonstrate that the specialist-interpolated DNN surrogate accurately reproduces the key flow and thermal features of the reference DSMC solution at an unseen Knudsen number.

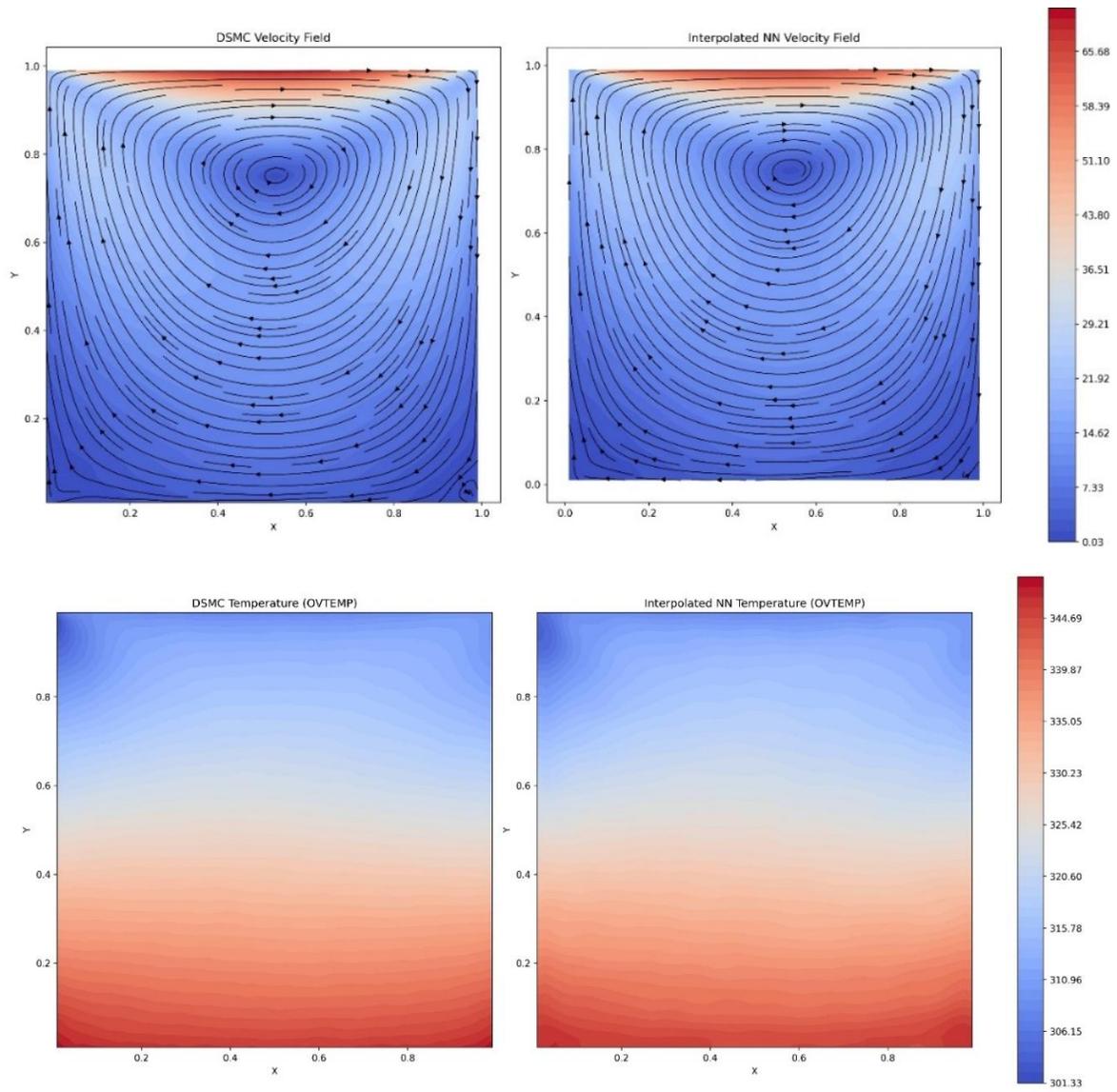

Fig. 14: DSMC and DNN prediction for the cavity flow velocity and temperature, Kn=0.05.

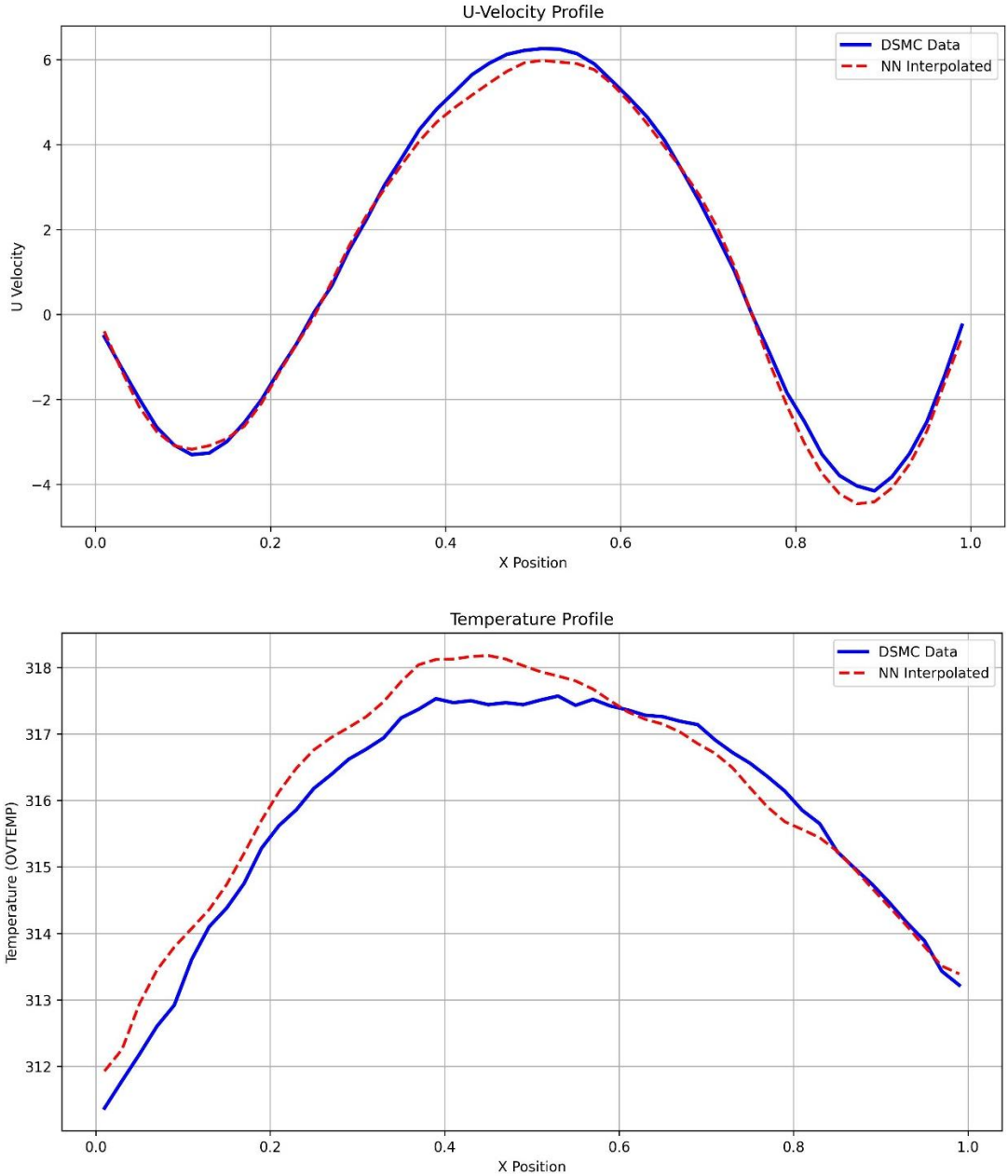

Fig. 15: Velocity and temperature distributions at Y/L=0.8 of the cavity predicted by DSMC and DNN.

Fig. 16 presents a comprehensive comparison between DSMC simulations and DNN predictions for lid-driven cavity flow at a Knudsen number of 0.5, representing a transitional rarefied flow regime. Both velocity fields exhibit the characteristic primary vortex near the center-top of the cavity, with very close alignment in magnitude and streamline structure, indicating strong agreement. The second row shows temperature contour plots, with the DSMC result on the left and the DNN result on the right. The DNN model accurately captures the temperature gradient from the bottom hot wall (350 K) to the cooler top wall (300 K), closely matching the reference DSMC solution, including fine thermal structures near the corners.

Fig. 17 shows quantitative comparisons along the vertical centerline (x = 0.5). The third panel presents the U-velocity profile, where the DNN prediction (dashed red) closely follows the DSMC result (solid blue), including the negative minimum near the bottom and sharp rise toward the top moving wall. The bottom panel displays the temperature profile along the same centerline, showing excellent agreement, which demonstrates that the DNN surrogate accurately captures both momentum and energy transport in this intermediate Knudsen-number flow. Confirming the outcome of recent literature [22-24], the figure validates the surrogate model's ability to predict high-fidelity flow fields in transitional rarefied regimes.

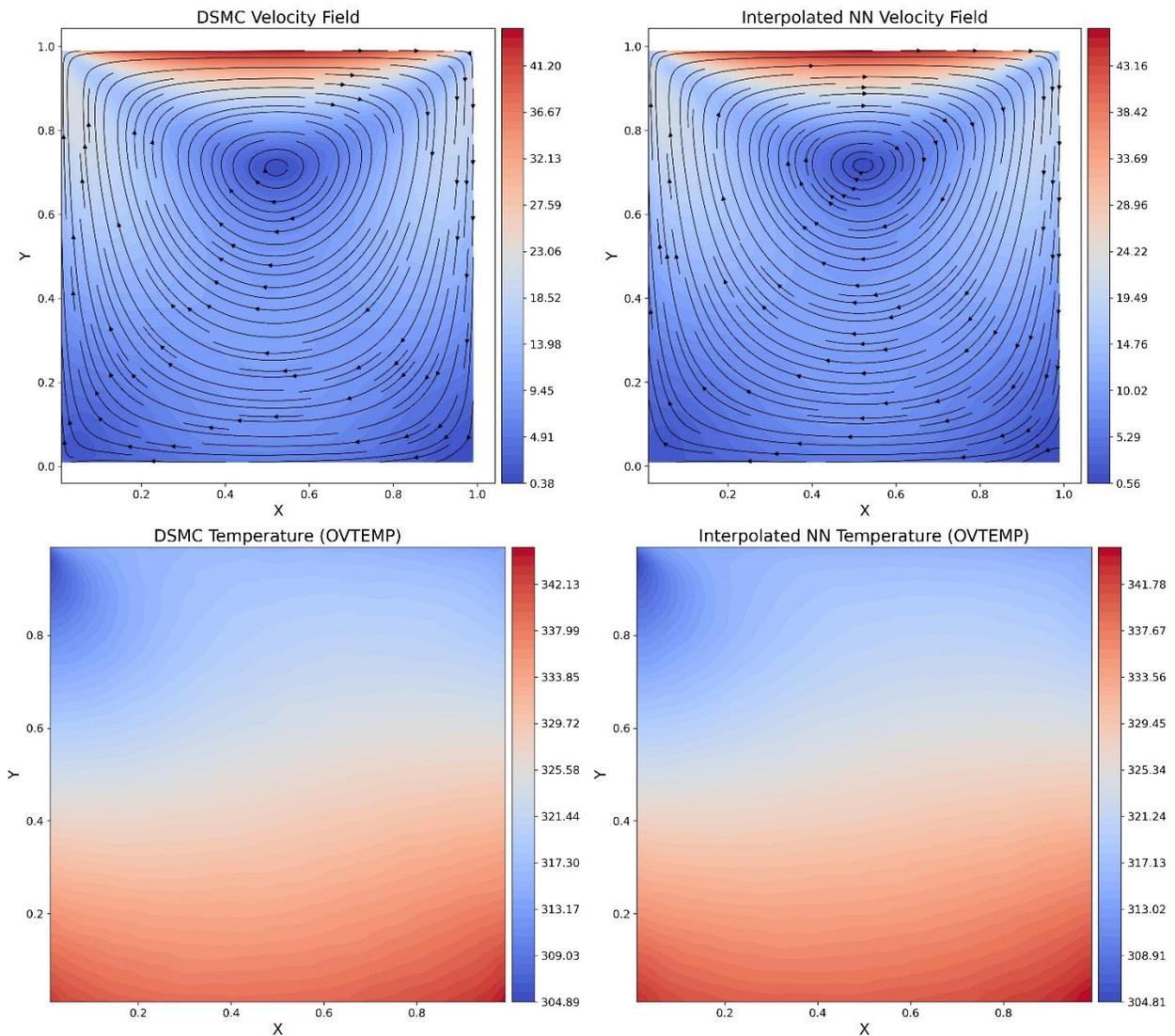

Fig. 16: DSMC and DNN contour field predictions for the cavity flow at Kn=0.5.

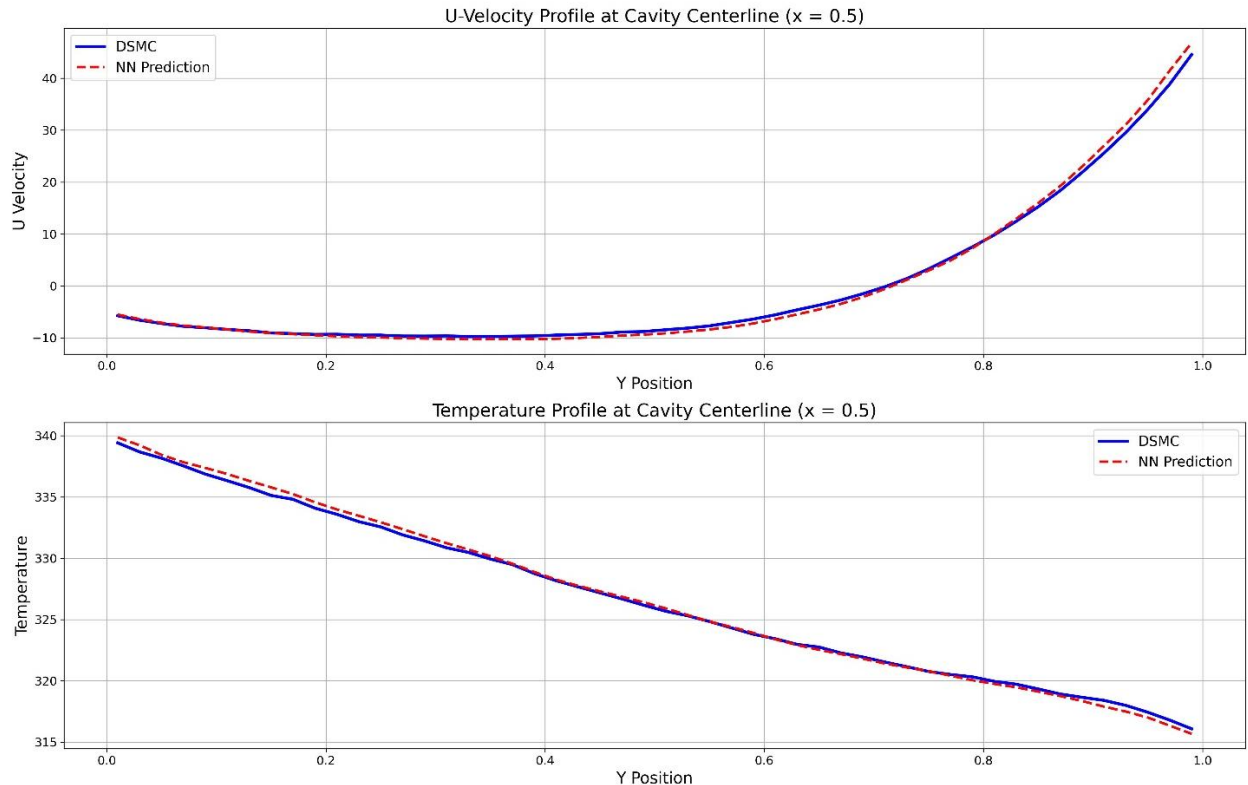

Fig. 17: DSMC and DNN detailed predictions for the cavity flow at Kn=0.5.

## 3. Concluding Remarks

In this work, we presented a robust and physically consistent surrogate modeling framework that leverages deep neural networks (DNNs) to accurately replicate the results of the DSMC simulations in rarefied gas dynamics. Our approach focuses on building high-fidelity neural surrogates that can emulate DSMC-computed flow fields, enabling orders-of-magnitude faster prediction while preserving physical accuracy. Using a family of specialist models trained individually on fixed Knudsen numbers, we decomposed the complex parametric dependence into simpler spatial learning tasks. This decomposition not only enhanced model stability but also facilitated the use of a log-space interpolation strategy to generalize to unseen intermediate Kn values.

The neural network architectures were carefully designed using Fourier feature mappings to handle high-frequency shock-like variations. They were regularized through dropout and L2 penalties to mitigate overfitting to the statistical noise inherent in DSMC data. The models demonstrated excellent agreement with DSMC across multiple flow regimes, from slip to transitional conditions, capturing both momentum and thermal transport with high accuracy. Importantly, the surrogates maintained predictive quality even under extrapolation, as demonstrated in Mach number generalization tasks, highlighting their reliability beyond the training distribution.

This study offers compelling evidence that deep learning, when combined with some physical insights and architectural regularization, can serve as an effective alternative to direct molecular simulations for many-query or real-time applications. The framework is particularly promising for accelerating parametric studies, uncertainty quantification, and design optimization in micro- and nano-scale gas flows. Future extensions will explore the incorporation of additional physical parameters, the expansion to three-dimensional configurations, and the integration of hybrid approaches with physics-informed neural networks (PINNs) to enhance generalizability and interpretability further.